\documentclass{svproc}
\usepackage{url}

\usepackage[algo2e]{algorithm2e} 
\usepackage{algorithm, algpseudocode}
\usepackage{color}
\usepackage{url}
\usepackage{hyperref}
\usepackage{amsmath, amssymb}
\usepackage{stackengine}
\usepackage{mathtools}
\usepackage{comment}
\usepackage[misc]{ifsym}
\usepackage{orcidlink}
\usepackage{graphicx}

\begin{document}
\mainmatter              % start of a contribution
\title{Classical and Quantum Random Walks to Identify Leaders in Criminal Networks}
\titlerunning{Classical and quantum random walks to identify leaders in criminal networks}  % abbreviated title (for running head)
%                                     also used for the TOC unless
%                                     \toctitle is used
%
\author{Annamaria Ficara\inst{1}$^{\textrm{(\Letter)}}$\orcidlink{0000-0001-9517-4131} \and Giacomo Fiumara\inst{1}\orcidlink{0000-0003-1528-7203} \and Pasquale De Meo\inst{2}\orcidlink{0000-0001-7421-216X} \and Salvatore Catanese\inst{1}\orcidlink{0000-0002-0369-8235}}
\authorrunning{Annamaria Ficara et al.} % abbreviated author list (for running head)
%
%%%% list of authors for the TOC (use if author list has to be modified)
\tocauthor{Annamaria Ficara, Giacomo Fiumara, Pasquale De Meo}
\institute{University of Messina, MIFT Department, 98166 Messina, Italy,\\
\email{\{aficara,gfiumara,scatanese\}@unime.it}
\and
University of Messina, DICAM Department, 98168 Messina, Italy, \\
\email{pdemeo@unime.it}}

\maketitle              % typeset the title of the contribution

\begin{abstract}
Random walks simulate the randomness of objects, and are key instruments in various fields such as computer science, biology and physics. The counter part of classical random walks in quantum mechanics are the quantum walks. Quantum walk algorithms provide an exponential speedup over classical algorithms. Classical and quantum random walks can be applied in social network analysis, and can be used to define specific centrality metrics in terms of node occupation on single-layer and multilayer networks. In this paper, we applied these new centrality measures to three real criminal networks derived from an anti-mafia operation named Montagna and a multilayer network derived from them. Our aim is to (i) identify leaders in our criminal networks, (ii) study the dependence between these centralities and the degree, (iii) compare the results obtained for the real multilayer criminal network with those of a synthetic multilayer network which replicates its structure.
\keywords{Classical random walks, quantum random walks, centrality, criminal networks}
\end{abstract}

\section{Introduction}

The term ``random walk'' was firstly introduced in 1905 by Pearson~\cite{pearson1905problem}. Random walks can be useful to solve practical problems. In fact, the randomness of objects can be analyzed and simulated through them. Also the correlation among objects can be computed using random walks. They are fast becoming key instruments in various fields such as physics, economics, computer science, biology, or chemistry~\cite{Xia2020Random}.

A simple random walk model in the mathematical space is a random process on a regular lattice where at each step one point jumps to another position according to a certain probability distribution~\cite{Xia2020Random}. Random walks can be applied to a network (or graph) defined as a set of nodes with edges between them. Graphs can be organized into multiple layers, which represent different types of nodes or edges. In this kind of application, the transition probability between nodes is greater when the strength of the association between nodes is stronger. After an appropriate number of steps, a random path able to describe the structure of a network can be obtained.
In computer science, the PageRank is the most known algorithm based on random walks~\cite{Paparo2013}. This algorithm randomly walks among web pages to compute their importance. 

Aharonov, Davidovich and Zagury were the first to propose quantum walks in 1993~\cite{aharonov1993quantum}. These walks can be considered as the counter part of classical random walks in quantum mechanics. The main difference between quantum walks and classical random walks is that the former do not converge to some limiting distributions~\cite{Xia2020Random}. Quantum interference can cause significantly faster or slower spreading of quantum walks compared to classical random walks. Algorithms based on quantum walks have lower time complexity with respect to the ones based on classical random walks, and they provide an exponential speedup over any classical algorithm~\cite{childs2003exponential}.

Random walks can find an application in computer vision, semi-supervised learning, network embedding, and complex social network analysis~\cite{li2015visual,Sarkar2011}. Some researchers also studied the application of random walks on knowledge discovery, graphs, science of science, and text analysis~\cite{Sarkar2011,Amancio2015Comparing,DEARRUDA2017154}. Quantum walks can be used instead to element distinctness, search problems, and decision trees~\cite{shenvi2003quantum,Ambainis2007Quantum}. Moreover, classical algorithms are often accelerated by using quantum walks.

Wald and B\"ottcher~\cite{Wald2021} at first introduced a framework for classical, quantum, and hybrid random walks with stochastic resetting on networks. Then, B\"ottcher and Porter~\cite{Bottcher2021} used continuous-time classical and quantum random walks to formulate occupation, PageRank, betweenness, and closeness centralities for multilayer networks. %To obtain relevant results beyond specific initial state and walk duration, they characterized a walk by the long-time average probability distribution of finding the walker at each node~\cite{Faccin2013Degree}, which captures all knowledge of the location of the walker in the absence of knowledge about when the walk began.

In this paper, we decided to apply classical and quantum occupation centralities on very specific kinds of real networks, \textit{i.e.} Mafia networks. These networks are different from traditional social networks because first of all they consist entirely of suspected criminals, and then they are considered as organizations of criminal groups and activities. For this reason, research on criminal networks is focused on their structure, efficiency, resilience and capacity to achieve criminal objectives.
To the best of our knowledge, classical and quantum walks have never used to identify leaders in criminal networks. Criminal leaders are strategic brokers within criminal networks~\cite{Calderoni2019}. Moreover, to reinforce our finding, we want to compare the results of classical and quantum walks on Mafia networks to the one obtained on Barab\'{a}si-Albert (BA)~\cite{Barabasi1999} models, which according to our previous work~\cite{Cavallaro2021}, are the one to better replicate criminal networks.

Mafia networks are in fact a good example of real-world networks with respect to the geometry of connections~\cite{FicaraCompleNet2021}. Indeed, these connections are the building blocks of the entire Mafia, and more generally of organized crime.
The Sicilian Mafia is one of the major native mafia-like organizations. It is composed by loose confederations of about one hundred families (or cosche). Since at least the 19th century, the social and economic life especially in Southern Italy is affected by these families~\cite{Paoli2008}.
Social network analysis and graph theory can be applied to Mafia networks to describe their structure and functioning, to identify leaders~\cite{eaifab2021}, to evaluate police interventions aimed at disrupting them~\cite{Cavallaro2020}, or to construct crime prevention systems~\cite{Calderoni2020}.

\section{Classical and Quantum Random Walk Centrality Measures}

\subsection{Classical Random Walks}

Random walks on graphs are sequences of nodes which start from a source node by selecting an edge, crossing the edge to reach a new node, and repeating the process. An equivalent concept can be found in the Markov chain which is a stochastic process that assumes values in a discrete set, where the next state of the chain only depends on the current state and not on the past states~\cite{portugal2013quantum}. 
The Markov chain can be viewed as a graph where nodes are represented by the states and edges by the possible next states, which are randomly determined. The set of states is discrete, but their evolution in terms of time can be discrete or continuous. 

In a discrete-time Markov chain, each step is associated with a probability distribution, \textit{i.e.} a set of probabilities that a walker is on a node or in a state. This probability distribution can be described with a vector, after choosing an order for the states. Given a graph $G(V,E)$ with set of nodes  $V = \{ v_1,\ldots, v_n\} \,(\lvert V\rvert = n)$ and set of edges $E$, the probability distribution is described by the vector $[p_1(t)\cdots p_n(t)]$, where $p_1(t)$ is the probability that a walker is on node $v_1$ at time $t$ and so on. If the process begins with the walker on the first node, $p_1(0) = 1$ and $p_i(0) = 0$ for $i=2,\ldots,n$. 

The future position of the walker cannot be precisely identified, but it is possible to determine a probability distribution knowing the transition matrix $M$.
By knowing the probability distribution at the time $t$, the distribution at time $t +1$ can be obtained by the following formula:
\begin{eqnarray}
p_i(t+1) = \sum_{j=1}^n M_{ij} p_j(t) \, .
\end{eqnarray}

In vector form, we have $\vec{p}(t+1) = M p(t)$.

The entry $M_{ij}$ of the transition matrix is the probability of the walker, who is on node $v_j$, to go to node $v_i$. The transition matrix is related to the adjacency matrix $A$ of the graph, and it can be defined as:
\begin{eqnarray}
M_{ij} = \frac{A_{ij} }{k_j} \, ,
\end{eqnarray}

where $k_j$ is the degree of node $v_j$, $A_{ij}=1$ when two nodes $v_i$ and $v_j$ are connected or $A_{ij}=0$ if they are not. If there is no edge from $v_j$ to $v_i$, the walker goes to one of the adjacent nodes and transition probability is the same for all of them. 

When time is a continuous variable, the problem of a walker who can go from node $v_j$ to a neighbour $v_i$ at any time can be seen as a liquid seeping from $v_j$ to $v_i$. At the beginning, it is likely that the walker is in the node $v_j$. As time goes by, the probability that the walker is in $v_j$ decreases, but the probability he is in one of its neighbors increases. The transition between neighbors occurs with a probability $\gamma$ per unit time, which is a transition rate presumably constant for all times and for all nodes. An infinitesimal time interval $\epsilon$ is generally used to set up and solve the differential equation of the problem with continuous variables.
Then, the probability that a walker goes from node $v_j$ to $v_i$ is given by $\gamma \epsilon$. If node $v_j$ has a degree $k_j$, it means it has $k_j$ neighbors. Therefore, the probability that the the walker is in one of the neighbors of $v_j$ after time $\epsilon$ is $k_j\gamma\epsilon$, while the probability of finding him in $v_j$ is $1-k_j\gamma\epsilon$. At this point, the entry $M_{ij}(t)$ of the transition matrix at time $t$ is defined as the probability that the walker on node $v_j$ goes to node $v_i$ in the time interval $t$:
\begin{eqnarray}
M_{ij}(\epsilon) = \begin{cases}
1 - k_j\gamma\epsilon+O(\epsilon^2), \quad \textrm{if} \ i=j;\\
\gamma\epsilon+O(\epsilon^2), \,\,\,\,\quad\qquad \textrm{if} \ i\neq j.
\end{cases}
\label{eq:Mij}
\end{eqnarray}

An auxiliary matrix called generating matrix $H$, is also defined as:
\begin{eqnarray}
H_{ij}(\epsilon) = \begin{cases}
k_j\gamma, \quad \textrm{if} \ i=j;\\
-\gamma, \,\quad \textrm{if} \ i\neq j \ \textrm{and adjacent};\\
0, \qquad \textrm{if} \ i\neq j \ \textrm{and non-adjacent.}
\end{cases}
\label{eq:Hij}
\end{eqnarray}

The next state of a Markov chain only depends on the current configuration of the chain. The transition matrix can be multiplied at different times as follows:
\begin{eqnarray}
M_{ij}(t+\epsilon) = \sum_k M_{ik}(t) M_{kj}(\epsilon) \, .
\end{eqnarray}

The index $k$ runs over all the neighbours of node $v_j$. In fact, if the walker is in node $v_j$, the probability that he goes to $v_k$ in the time interval is $M_{kj}(\epsilon)$, independently of the value of $k$. If there is no edge between $v_j$ and $v_k$ for a specific $k$, $M_{kj}(\epsilon)=0$.

By isolating the term $k=j$, using the Eqs.~\ref{eq:Mij} and~\ref{eq:Hij}, and then by moving the first term on the right-hand side to the left-hand side and dividing it by $\epsilon$, the following differential equation is obtained:
\begin{eqnarray}
\frac{d M_{ij}(t)}{dt} = - \sum_{k} H_{kj} M_{ik}(t) \, 
\label{eq:df}
\end{eqnarray}

The solution of Eq.~\ref{eq:df} with initial condition $M_{ij}(0) = \delta_{ij}$ is:
\begin{eqnarray}
M(t) = e^{-Ht} \, .
\label{eq:Mt}
\end{eqnarray}

%The verification is simple, if we expand the exponential function in Taylor series. 
After the definition of the transition matrix, the probability distribution at time $t$ can be easily obtained. If the initial distribution is $\vec{p}(0)$, $\vec{p}(t) = M(t)\vec{p}(0)$.

\subsection{Quantum Random Walks}

The electron spin is described by quantum mechanics as a unit vector in the Hilbert space $\mathbb{C}^2$~\cite{portugal2013quantum}. The spin up and the spin down are respectively described by the vectors $\lvert 0\rangle$ and $\lvert 1\rangle$: 
\begin{eqnarray}
\setstackgap{L}{1.2\baselineskip}
\lvert 0\rangle = \bracketVectorstack{1 0} \, , \quad \lvert 1\rangle = \bracketVectorstack{0 1} \, .
\end{eqnarray}

The notions of spin up and spin down refer to $\mathbb{R}^3$. Quantum mechanics describes in fact the behavior of the electron before entering the magnetic field. In this case and if the electron is somehow isolated from the macroscopic environment, its spin state is described by a linear combination of vectors $\lvert 0\rangle$ and $\lvert 1\rangle$. Therefore, $\lvert \psi\rangle = a_0\lvert 0\rangle + a_1\lvert 1\rangle$, where the coefficients $a_0$ and $a_1$ are complex numbers which satisfy the constraint $\lvert a_0\rvert^2 + \lvert a_1\rvert^2 = 1$.

The time evolution of an isolated quantum system is described by a unitary transformation. If the state of the quantum system at time $t_1$ is described by vector $\lvert \psi_1\rangle$, the system state $\lvert \psi_2\rangle$ at time $t_2$ is obtained from $\lvert \psi_1\rangle$ by a unitary transformation $U$, which depends only on $t_1$ and $t_2$. Therefore, we have $\lvert \psi_2\rangle = U\lvert \psi_1\rangle$.

The evolution postulate is to be written in the form of a differential equation which provides a method to obtain operator $U$ once given the physical context. It is called Schr\"odinger equation and it is as follows:
\begin{eqnarray}
\frac{d \lvert\psi\rangle}{dt} = - iH\lvert\psi\rangle \, .
\label{eq:schrodinger}
\end{eqnarray}

In the passage from the continuous-time classical walk model to the continuous-time quantum walk model, the transition matrix is simply converted to an equivalent unitary operator, and the vector which describes the probability distribution to a state vector. Matrix $H$, given by Eq.~\ref{eq:Hij} is Hermitian, and therefore matrix $M$ given by Eq.~\ref{eq:Mt} is not unitary. $M$ can be easily made unitary within the context of Hilbert spaces by multipling $h$ with the imaginary unit, that is by replacing $H$ with $iH$. Then, the evolution operator of the continuous-time quantum walk can be defined as follows:
\begin{eqnarray}
U(t) = e^{-iHt} \, .
\label{eq:evolop}
\end{eqnarray}

If the initial condition is $\lvert \psi(0)\rangle$, the quantum state at time $t$ is:
\begin{eqnarray}
\lvert \psi(t)\rangle = U(t) \lvert \psi(0)\rangle \, ,
\end{eqnarray}

and the probability distribution is:
\begin{eqnarray}
p_k = \lvert \langle k \lvert \psi(t)\rangle \rvert^2 \, ,
\end{eqnarray}

where $k$ runs over all nodes of the graph (or states of the Markov chain) and $\lvert k\rangle$ is the state of the computational basis which correspond to the node $v_k$. The computational basis of space $\mathbb{C}^2$ is the set $\{\lvert 0\rangle,\lvert 1\rangle\}$.

\subsection{Random Walk Occupation Centralities}
\label{subsec:centrality}

Recently continuous-time classical and quantum random walk centrality measures was defined for single-layer and multilayer networks~\cite{Wald2021,Bottcher2021}.

Starting from Eq.~\ref{eq:Mt}, the explicit Euler integration scheme can be used to simulate a classical random walker:
\begin{eqnarray}
p_{t+1} = 1 - H_c \, \Delta t \, p_t  \, ,
\end{eqnarray}
where $H_c$ is the classical Hamiltonian defined as $H_c = LD^{-1}$, where $L=D-A$ is the Laplacian matrix and $D$ is the degree matrix with $D_{ii} = k_i$ if $i=j$ and $D_{ij} = 0$ if $i\neq j$. 

The classical random walks approaches for connected networks and sufficiently long times a stationary probability distribution $p^*$ such that:
\begin{eqnarray}
p^*_i = \sum\limits_{j=1}^n\frac{A_{ij}}{k_j} p^*_j = \frac{k_i}{\sum\limits_{i=1}^{n}k_i}  \, .
\label{eq:opc}
\end{eqnarray}

The stochastic vector $p^*_i$ indicates the stationary classic random walk occupation probability $OP_c$ for a node $v_i$, and it can be used to define a random walk centrality measure for graphs. 

The Schr\"odinger equation (see Eq.~\ref{eq:schrodinger}) can be instead solved using the Crank-Nicholson integration scheme~\cite{Askar1978,Wald2021}. Expanding the evolution operator in Eq.~\ref{eq:evolop} into a Taylor series, we have:
\begin{eqnarray}
\lvert\psi_{t+1}\rangle = -2i\Delta t H_q \lvert\psi_t\rangle+\lvert\psi_{t-1}\rangle \, ,
\end{eqnarray}
where $H_q$ is the Hermitian quantum Hamiltonian defined as $H_q = \mathcal{L}$, where $ \mathcal{L}= D^{-\frac{1}{2}} LD^{-\frac{1}{2}}$ is the normalized version of the Laplacian matrix $\mathcal{L}$.

For continuous-time quantum walks, unitary time evolution does not lead to a stationary state. Instead, the long-time behavior of a quantum walk is characterized by its long-time mean:
\begin{eqnarray}
q^*_i = \lim\limits_{T\to\infty}\frac{1}{T}\int_{0}^{T} \langle i\rvert\lvert\psi(t)\rangle\langle\psi(t)\rvert \lvert i\rangle dt
\label{eq:opq}
\end{eqnarray}

where $dt$ is an infinitesimal time step, $\lvert\psi(t)\rangle\langle\psi(t)\rvert$ is a density operator, and $\lvert i\rangle \in \mathbb{C}^\mathcal{N}$ is an orthonormal basis vector. 

Even in this case, $q^*_i$ denotes the continuous-time quantum random walk occupation probability $OP_q$ of a node $v_i$, and therefore it gives a type of random walk centrality.

\section{Methods and Results}

\begin{figure}[ht!]
\centering
\includegraphics[width=\textwidth]{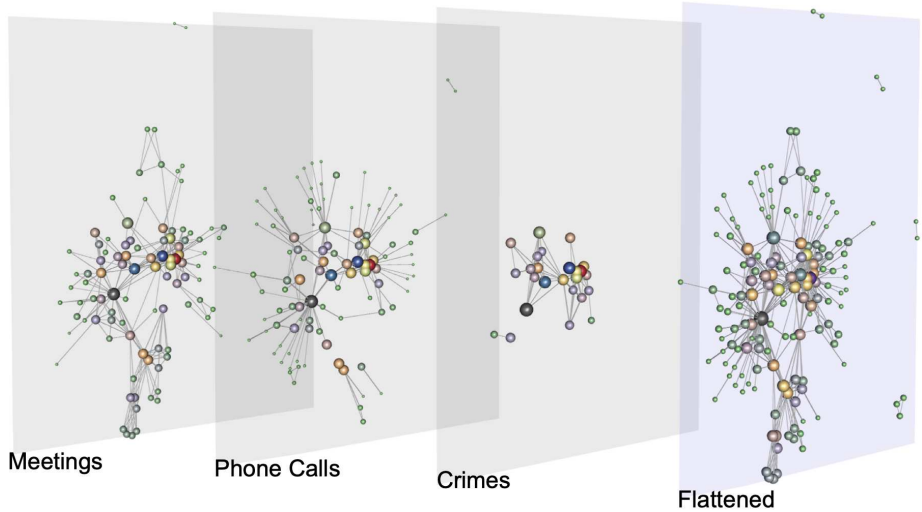}%scale=0.15
\caption{Meetings, Phone Calls, Crimes and the flattened multilayer network from the Montagna anti-mafia operation.}
\label{fig:fig1}
\end{figure}

We used the classical and quantum occupation centralities described in Subsect.~\ref{subsec:centrality} to study the importance of nodes in three real single-layer criminal networks (Meetings, Phone Calls and Crimes) related to an anti-mafia operation called Montagna~\cite{Calderoni2020,Cavallaro2020,ficara2021criminal}, from which we derived a fourth network in the form of a flattened multilayer network. The Montagna operation focused on the Mistretta and Batanesi clans who monopolized the sector of public contracts in the Tyrrhenian strip and in the nebroidal district of the province of Messina, by working with some entrepreneurs associated to the Sicilian Mafia. Meetings contains $256$ physical meetings between $101$ suspected criminals observed during the police physical surveillance. Phone Calls contains $124$ phone calls between $100$ suspects wiretapped during the police audio surveillance. Crimes contains $74$ connections between $25$ suspected criminals who committed crimes together. The three networks share $20$ nodes. We also created an undirected and weighted multilayer network with $226$ actors, $454$ intralayer edges and $3$ layers corresponding to the three single-layer criminal networks Meetings, Phone Calls and Crimes.  
In this work, a single-layer network (flattened) is created by merging the three layers of the multilayer network. In the flattened network, there is one node for each actor of the starting multilayer network and an edge between two nodes if the corresponding actors are linked in one layer. 
Meeting, wiretap and co-offending networks provide fundamentally different pictures of the criminal organization, and therefore are all particularly useful in finding leaders. Also the flattened multilayer network can be useful to have a more complete picture of the organization.
Figure~\ref{fig:fig1} shows Meetings, Phone Calls, Crimes and the flattened multilayer network.

Our analysis was conducted using the mathematical framework\footnote{Available at: \url{https://gitlab.com/ComputationalScience/ctqw_centrality}} of continuous-time classical and quantum centrality measures created by B\"ottcher and Porter~\cite{Bottcher2021}. %which is applicable to both single-layer and multilayer networks~\cite{Bottcher2021}.

\begin{figure}[ht!]
\centering
\includegraphics[width=\textwidth]{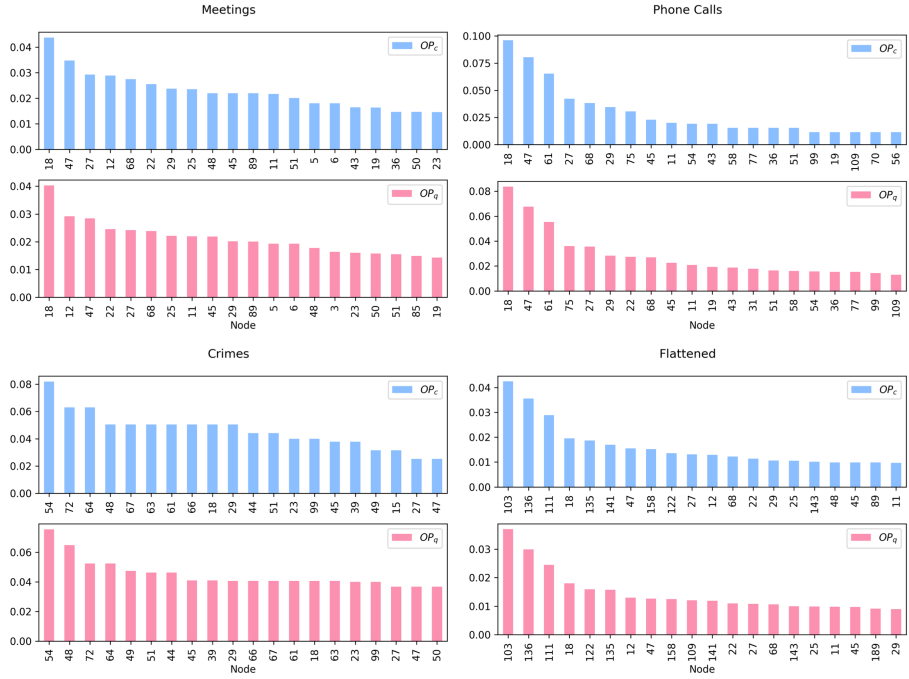}
\caption{20 top ranked nodes computed with classical $OP_c$ and quantum $OP_q$ random walk occupation probabilities in the single layer networks Meetings, Phone Calls and Crimes, and the flattened Montagna multilayer network.}
\label{fig:fig3}
\end{figure}

The top $20$ nodes ranked by their classical and quantum occupation probabilities are shown in Fig.~\ref{fig:fig3}. In the analysis of central nodes in a Mafia network, we have to consider the hierarchical structure of a Mafia family in which individuals can be categorized as boss, underboss, consigliere (\textit{i.e.}, an advisor to the boss), messaggero (\textit{i.e.}, a messenger among families), caporegime (\textit{i.e.}, who heads a local crew of soldiers), soldier (\textit{i.e.}, average type criminal) and associate (\textit{e.g.}, entrepreneurs, pharmacists, police officers, politician etc.).
The results for the Meetings and Phone Calls networks confirm the one obtained through the application of the degree as we did in~\cite{eaifab2021}, revealing that the most important actors are nodes $18$ and $47$. These nodes are respectively Caporegime of the Mistretta family and deputy Caporegime of the Batanesi family. The application of $OP_C$ and $OP_q$ on the Crimes network reveals instead different central nodes who are not key members of the Mafia hierarchy but are associates such as entrepreneurs (\textit{i.e.}, nodes $54$ and $64$), construction workers (\textit{i.e.}, node $72$) or soldiers (\textit{i.e.}, nodes $48$ and $49$). Differently from the application of degree on the multilayer network, in this case, we obtain different central nodes such as $103$, $136$ and $111$. Nodes $103$ and $111$ were intercepted because they were in phone contact respectively with a Caporegime of the Mistretta family (\textit{i.e.}, node $18$) and a non-existent entrepreneur, created for the purpose of fraud (\textit{i.e.}, node $109$). Node $136$ is instead a cohabitant of a Caporegime of the Batanesi family (\textit{i.e.}, node $27$). The role of these figures in the mafia business in unclear but their centrality can direct investigations and lead to deeper investigation of them. 
Classical and quantum occupation probabilities identify in most cases the same leaders in criminal networks. Therefore, when a micro approach is adopted in SNA application to organized crime focusing, as in our case, on one organization or network with a relatively low number of nodes, there is no clear advantage to using quantum random walks instead of classical ones. However, quantum walks can be useful to speed up the investigations, and to find leaders when a macro approach is adopted to study crime through SNA. In this case, in fact, larger networks focusing on specific national or regional criminal markets or offender categories are analyzed.

\begin{figure}[ht!]
\centering
\includegraphics[width=\textwidth]{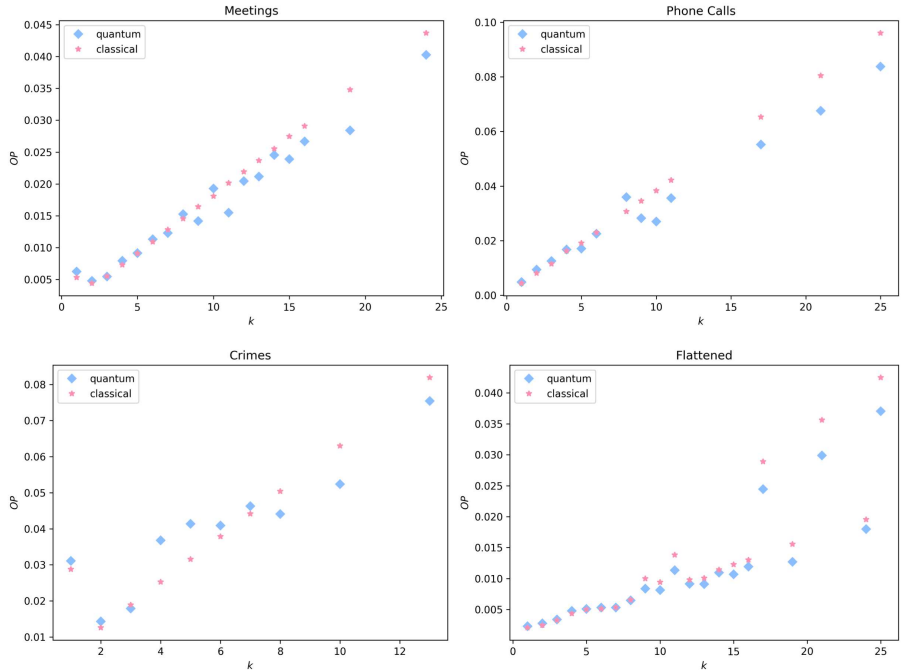}
\caption{Classical and quantum random walk occupation probabilities for the single layer networks Meetings, Phone Calls and Crimes, and the flattened Montagna multilayer network that a node of degree $k$ is occupied by a classical and quantum random walker, respectively.}
\label{fig:fig2}
\end{figure}

In Fig.~\ref{fig:fig2}, we show the probabilities that a node with a certain degree is occupied by a classical and quantum random walker. As explained in~\cite{Bottcher2021}, we can observe a linear dependence between $OP_c$ and $OP_q$ which is due to the fact that quantum walks satisfy Eq.~\ref{eq:opc} because they do not approach a stationary state. Their long-time behavior is instead characterized by the long-time mean as shown in Eq.~\ref{eq:opq}. Quantum walks are not proportional to the degree which is not able to capture node occupation properties of this kind of walks. These properties, in fact, also depend on other structural features of the underlying networks. The minimum and maximum degrees of the Meetings, Phone Calls, Crimes networks are respectively $1$ and $24$, $1$ and $25$, $1$ and $13$, as shown in the plot of the degree distribution in~\cite{fi14050123}. The differences between these classical and quantum centralities seem to be greater in the flattened multilayer network.

\begin{figure}[ht!]
\centering
\includegraphics[width=\textwidth]{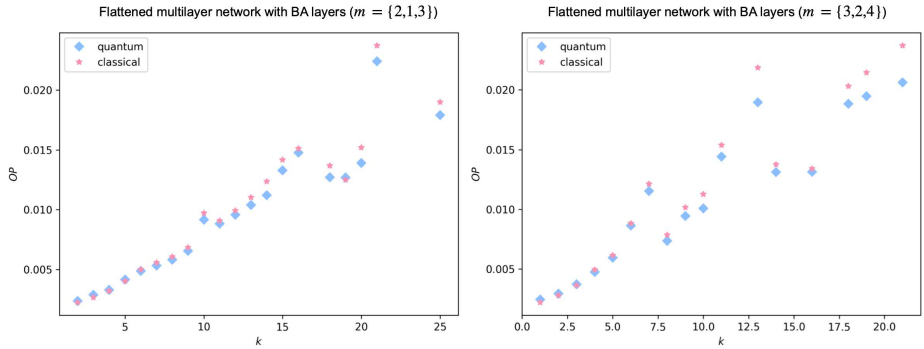}
\caption{Classical and quantum random walk occupation probabilities for a flattened multilayer network composed by three Barab\'{a}si-Albert layers with $101$ nodes and $198$ edges, $100$ nodes and $99$ edges, $25$ nodes and $66$ edges (left column); $101$ nodes and $294$ edges, $100$ nodes and $196$ edges, $25$ nodes and $84$ edges (right column).}
\label{fig:fig4}
\end{figure}

In~\cite{Bottcher2021}, the authors asserted that the differences between the classical and quantum occupation centralities were (i) larger in BA single-layer networks~\cite{Wald2021} or in synthetic multilayer networks which include BA layers, and (ii) smaller than those observed for synthetic networks.
The second results is not valuable for real criminal networks, and this can be explained just through the first finding. In our previous work~\cite{Cavallaro2021}, we used some popular network models such as Erd\"{o}s-R\'{e}nyi, Watts-Strogatz and different configurations of BA to replicate the topology of a criminal network. Our experiments identified the BA model as the one which better represents a criminal network. 
To prove our thesis, we decided to build a flattened multilayer network with three BA layers. Each layer is created with the same number of nodes of the Meetings, Phone Calls, Crimes networks (\textit{i.e.}, $101$, $100$, and $25$, respectively) and a number of edges as close as possible to that of these networks (\textit{i.e.}, $256$, $124$, and $74$, respectively). BA models are built based on a specific parameter $m$, which indicates the number of edges that are preferentially attached to existing nodes with high degree. For the first layer, we tried two different configurations of the BA graphs with $m=2$ (obtaining $198$ edges) and $m=3$ (obtaining $294$ edges). For the second layer, we used $m=1$ (obtaining $99$ edges) and $m=2$ (obtaining $196$ edges). For the third layer, we choose $m=3$ (obtaining $66$ edges) and $m=4$ (obtaining $84$ edges).
Figure~\ref{fig:fig4} shows classical and quantum occupation centralities versus the node degrees on the described synthetic multilayer network.

\section{Conclusions and Discussion}
In this paper, we used classical and quantum continuous-time random walk centrality measures to identify leaders in (i) three single-layer criminal networks related to an anti-Mafia operation called Montagna, and (ii) a flattened multilayer network created from them. 

In the formulation of these measures, different Hamiltonians describe the evolution of classical and quantum walks. 
Other classical and quantum continuous-time random walk centralities, such as random walk betweenness or closeness, can be derived by modifying these Hamiltonians, and tracking different properties of absorbing random walks.
In our future works, we want to apply these other measures on our criminal networks to identify leaders, and to define a new disruption framework based on random walks, comparing the targeted attacks based on the classical centrality measures with those based on their random walk versions.

Moreover, we are planning to extend the mathematical framework by B\"ottcher and Porter~\cite{Bottcher2021} for whole multilayer networks considering the layered structure, and not just the flattening function. In fact, we can obtain a complete picture of the activities and the structure of criminal organizations only considering the various layers not as single-layer networks but as a whole multilayer network~\cite{fi14050123}. Then, we want to explore the possibility of developing a disruption framework for multilayer networks.

%
% ---- Bibliography ----
%
\bibliographystyle{splncs03}
\bibliography{bibfile}

\begin{thebibliography}{10}
\providecommand{\url}[1]{\texttt{#1}}
\providecommand{\urlprefix}{URL }

\bibitem{aharonov1993quantum}
Aharonov, Y., Davidovich, L., Zagury, N.: Quantum random walks. Physical Review
  A  48(2),  1687 (1993)

\bibitem{Amancio2015Comparing}
Amancio, D.R.: Comparing the topological properties of real and artificially
  generated scientific manuscripts. Scientometrics  105(3),  1763--1779 (2015)

\bibitem{Ambainis2007Quantum}
Ambainis, A.: Quantum walk algorithm for element distinctness. SIAM Journal on
  Computing  37(1),  210--239 (2007)

\bibitem{Askar1978}
Askar, A., Cakmak, A.S.: Explicit integration method for the time‐dependent
  schrodinger equation for collision problems. The Journal of Chemical Physics
  68(6),  2794--2798 (1978)

\bibitem{Barabasi1999}
Barab{\'{a}}si, A.L., Albert, R.: {Emergence of Scaling in Random Networks}.
  Science  286(5439),  509--512 (oct 1999)

\bibitem{Bottcher2021}
B\"ottcher, L., Porter, M.A.: Classical and quantum random-walk centrality
  measures in multilayer networks. SIAM Journal on Applied Mathematics  81(6),
  2704--2724 (2021)

\bibitem{Calderoni2020}
Calderoni, F., Catanese, S., {De Meo}, P., Ficara, A., Fiumara, G.: Robust link
  prediction in criminal networks: A case study of the sicilian mafia. Expert
  Systems with Applications  161,  113666 (2020)

\bibitem{Calderoni2019}
Calderoni, F., Superchi, E.: The nature of organized crime leadership: criminal
  leaders in meeting and wiretap networks. Crime, Law and Social Change  72(4),
   419--444 (2019)

\bibitem{Cavallaro2021}
Cavallaro, L., Ficara, A., Curreri, F., Fiumara, G., De~Meo, P., Bagdasar, O.,
  Liotta, A.: Graph comparison and artificial models for simulating real
  criminal networks. In: Benito, R.M., Cherifi, C., Cherifi, H., Moro, E.,
  Rocha, L.M., Sales-Pardo, M. (eds.) Complex Networks and Their Applications
  IX. pp. 286--297. Springer, Cham (2021)

\bibitem{Cavallaro2020}
Cavallaro, L., Ficara, A., {De Meo}, P., Fiumara, G., Catanese, S., Bagdasar,
  O., Song, W., Liotta, A.: Disrupting resilient criminal networks through data
  analysis: The case of {Sicilian Mafia}. PLOS ONE  15(8),  1--22 (2020)

\bibitem{childs2003exponential}
Childs, A.M., Cleve, R., Deotto, E., Farhi, E., Gutmann, S., Spielman, D.A.:
  Exponential algorithmic speedup by a quantum walk. In: Proceedings of the
  thirty-fifth annual ACM symposium on Theory of computing. pp. 59--68 (2003)

\bibitem{DEARRUDA2017154}
{de Arruda}, H.F., Silva, F.N., da~F.~Costa, L., Amancio, D.R.: Knowledge
  acquisition: A complex networks approach. Information Sciences  421,
  154--166 (2017)

\bibitem{ficara2021criminal}
Ficara, A., Cavallaro, L., Curreri, F., Fiumara, G., De~Meo, P., Bagdasar, O.,
  Song, W., Liotta, A.: Criminal networks analysis in missing data scenarios
  through graph distances. PLOS ONE  16(8),  1--18 (2021)

\bibitem{fi14050123}
Ficara, A., Fiumara, G., Catanese, S., De~Meo, P., Liu, X.: The whole is
  greater than the sum of the parts: A multilayer approach on criminal
  networks. Future Internet  14(5),  123 (2022)

\bibitem{eaifab2021}
Ficara, A., Fiumara, G., De~Meo, P., Catanese, S.: Multilayer network analysis:
  The identification of key actors in a sicilian mafia operation. In:
  Perakovic, D., Knapcikova, L. (eds.) Future Access Enablers for Ubiquitous
  and Intelligent Infrastructures. pp. 120--134. Springer, Cham (2021)

\bibitem{FicaraCompleNet2021}
Ficara, A., Saitta, R., Fiumara, G., De~Meo, P., Liotta, A.: Game of thieves
  and werw-kpath: Two novel measures of node and edge centrality for mafia
  networks. In: Teixeira, A., Pacheco, D., Oliveira, M., Barbosa, H.,
  Gon{\c{c}}alves, B., Menezes, R. (eds.) Complex Networks XII. pp. 12--23.
  Springer, Cham (2021)

\bibitem{li2015visual}
Li, X., Han, Z., Wang, L., Lu, H.: Visual tracking via random walks on graph
  model. IEEE transactions on Cybernetics  46(9),  2144--2155 (2015)

\bibitem{Paoli2008}
Paoli, L.: {Mafia brotherhoods: Organized crime, Italian style}. OUP (2008)

\bibitem{Paparo2013}
Paparo, G.D., M{\"u}ller, M., Comellas, F., Martin-Delgado, M.A.: Quantum
  google in a complex network. Scientific Reports  3(1),  2773 (2013)

\bibitem{pearson1905problem}
Pearson, K.: The problem of the random walk. Nature  72(1867),  342--342 (1905)

\bibitem{portugal2013quantum}
Portugal, R.: Quantum Walks and Search Algorithms. Quantum Science and
  Technology, Springer New York (2013)

\bibitem{Sarkar2011}
Sarkar, P., Moore, A.W.: Random walks in social networks and their
  applications: A survey. In: Aggarwal, C.C. (ed.) Social Network Data
  Analytics. pp. 43--77. Springer US, Boston, MA (2011)

\bibitem{shenvi2003quantum}
Shenvi, N., Kempe, J., Whaley, K.B.: Quantum random-walk search algorithm.
  Physical Review A  67(5),  052307 (2003)

\bibitem{Wald2021}
Wald, S., B\"ottcher, L.: From classical to quantum walks with stochastic
  resetting on networks. Phys. Rev. E  103,  012122 (2021)

\bibitem{Xia2020Random}
Xia, F., Liu, J., Nie, H., Fu, Y., Wan, L., Kong, X.: Random walks: A review of
  algorithms and applications. IEEE Transactions on Emerging Topics in
  Computational Intelligence  4(2),  95--107 (2020)

\end{thebibliography}

\end{document}